\documentclass[conference]{IEEEtran}
\usepackage[pdftex]{graphicx}
\usepackage{amsmath}
\usepackage{amssymb}
\graphicspath{{images/}}

\begin{document}
\title{An Ultra-Fast Method for Simulation of Realistic Ultrasound Images}

\author{\
\IEEEauthorblockN{Mostafa~Sharifzadeh, Habib~Benali, and Hassan~Rivaz}
\IEEEauthorblockA{Department of Electrical and Computer Engineering\\Concordia University\\ Montreal, QC, Canada\\mostafa.sharifzadeh@concordia.ca}}
\maketitle

\begin{abstract}
Convolutional neural networks (CNNs) have attracted a rapidly growing interest in a variety of different processing tasks in the medical ultrasound community. However, the performance of CNNs is highly reliant on both the amount and fidelity of the training data. Therefore, scarce data is almost always a concern, particularly in the medical field, where clinical data is not easily accessible. The utilization of synthetic data is a popular approach to address this challenge. However, simulating a large number of images using packages such as Field II is time-consuming, and the distribution of simulated images is far from that of the real images. Herein, we introduce a novel ultra-fast ultrasound image simulation method based on the Fourier transform and evaluate its performance in a lesion segmentation task. We demonstrate that data augmentation using the images generated by the proposed method substantially outperforms Field II in terms of Dice similarity coefficient, while the simulation is almost 36000 times faster (both on CPU).
\end{abstract}

\IEEEpeerreviewmaketitle

\section{Introduction}
In the domain of medical ultrasound (US) image analysis, approaches based on convolutional neural networks (CNNs) have gained significant attention over the past few years as they outperformed classical methods in a variety of applications. However, in these approaches, there is always a trade-off between the performance and available data, and CNNs with high learning capacities often suffer from the overfitting problem due to scarce data in the medical field. Since augmenting the dataset using synthetic data is a common solution to address this problem, data simulation has become a critical task and its importance has been heightened more than ever before.

Simulating US images have been extensively investigated in the medical context, and several publicly available packages have been released for this purpose \cite{focuswebsite, Treeby2010, Jensen1996}. Solving acoustic wave equations in the medium is one of the most well-known approaches to that aim \cite{Verweij2014}, where complex equations make it computationally expensive and relatively slow.
Treeby \textit{et al.} used the \textit{k}-space pseudospectral method to reduce the complexity for modeling nonlinear US propagation in heterogeneous media with power law absorption \cite{Treeby2012a}.
Jensen \textit{et al.} suggested calculating pulsed pressure fields based on the Tupholme-Stepanishen method, wherein shape, excitation, and apodization of the transducer could be set as parameters. They divided the surface into small rectangular patches and calculated the field in each one to obtain the final field by summing their responses. Besides, they used a far-field approximation instead of the geometric one in favor of a faster calculation compared to the older methods \cite{Jensen1992}.

Another approach for US simulation is based on ray-tracing methods,  where the graphics processing unit (GPU) is employed to simulate the propagation of the US wavefront as rays. Given scatterers' distribution, this approach is capable of generating the speckle pattern by convolving a point spread function (PSF) with scatterers, while simulating complex US interactions, such as refractions and reflections.
Bürger \textit{et al.} suggested a simulation method based on a convolution-enhanced ray-tracing approach and employed a deformable mesh model. They demonstrated that a better simulation of artifacts is achievable by following the path of the US pulse \cite{Burger2013}.
Mattausch \textit{et al.} proposed using interactive Monte-Carlo path tracing for simulation of complex surface interactions, which enables more realistic simulation of tissue interactions, such as soft shadows and fuzzy reﬂections \cite{Mattausch2018}.

Recently, deep learning approaches are also exploited for synthesizing US images.
Zhang \textit{et al.} demonstrated an approach to estimate the probabilistic scatterer from observed US data by imposing a known statistical distribution on scatterers and learn the mapping between US image and distribution parameter map by training a CNN on synthetic images \cite{Zhang2020}.
Hu \textit{et al.} proposed a method based on a conditional generative adversarial network (GAN) to simulate US images at given 3D spatial locations relative to the patient anatomy \cite{Hu2017}.
Cronin \textit{et al.} investigated a framework that accepts synthetic masks and real images as inputs of a GAN and generates realistic B-mode musculoskeletal US images that are statistically similar to real images \cite{Cronin2020}.
Liang \textit{et al.} introduced an end-to-end framework for enhancing the structure fidelity and resolution of simulated images by employing a sketch GAN and a progressive training strategy and validated that on the follicle and ovary US image synthesis \cite{Liang2020}.
GANs were also adopted for simulating intraoperative US images of the brain after tumor resection surgery \cite{Donnez2021}, intravascular \cite{Tom2018}, and kidney US images \cite{Pigeau2020}.

In this work, we propose a paradigm shift for ultra-fast simulation of B-mode US images. Our approach is entirely different from the abovementioned methods and is based on the Fourier transform. Besides, we adopt this method for augmenting the training set using simulated images in a lesion segmentation task and demonstrate a substantial improvement in the performance.

\section{Ultra-fast simulation}
To simulate a new US image containing lesion(s) with known ground truth, we propose taking a real US image and an arbitrary mask (as the ground truth) and substituting the phase information of the low-frequency spectrum of the real image with the corresponding information of the mask.

Let $I_r$, $I_m$, and $I_s \in \mathbb{R}^{W\times H}$ represent a real US image, an arbitrary mask, and the new simulated output image, respectively.
In addition, let $\mathcal{F}_M(I): \mathbb{R}^{W\times H} \rightarrow \mathbb{R}^{W\times H}$ and $\mathcal{F}_P(I): \mathbb{R}^{W\times H} \rightarrow \mathbb{R}^{W\times H}$ denote the magnitude and phase of the Fourier transform $\mathcal{F}$ of the image $I$:
\begin{flalign}
\mathcal{F}(I)(m,n)=\sum_{w=0}^{W-1}\sum_{h=0}^{H-1}I(w,h)e^{\textstyle -j2\pi(\frac{h}{H}n+\frac{w}{W}m)}
\label{eq1}
\end{flalign}

\noindent Accordingly, given $\mathcal{F}_M(I)$ and $\mathcal{F}_P(I)$, $\mathcal{F}^{-1}$ is the inverse Fourier transform that converts back the signal from the frequency domain to the image domain.
\begin{flalign}
I = \mathcal{F}^{-1}(\mathcal{F}_M(I), \mathcal{F}_P(I))
\label{eq2}
\end{flalign}

\noindent where $j^2=-1$, and (\ref{eq1}) and (\ref{eq2}) can be implemented using FFT \cite{Frigo1998} and IFFT algorithms, respectively.

\noindent Further, let denote with $M_\alpha$ a matrix of size $W \times H$:
\begin{flalign}
M_\alpha(w,h) =
\begin{cases} 
1, & \frac{(w-\frac{W}{2})^2}{(\alpha \frac{W}{2})^2}+\frac{(h-\frac{H}{2})^2}{(\alpha \frac{H}{2})^2}\leq 1\\
0, & otherwise
\end{cases}
\label{eq3}
\end{flalign}

\begin{figure*}
	\centering
	\includegraphics[width=0.9999\linewidth]{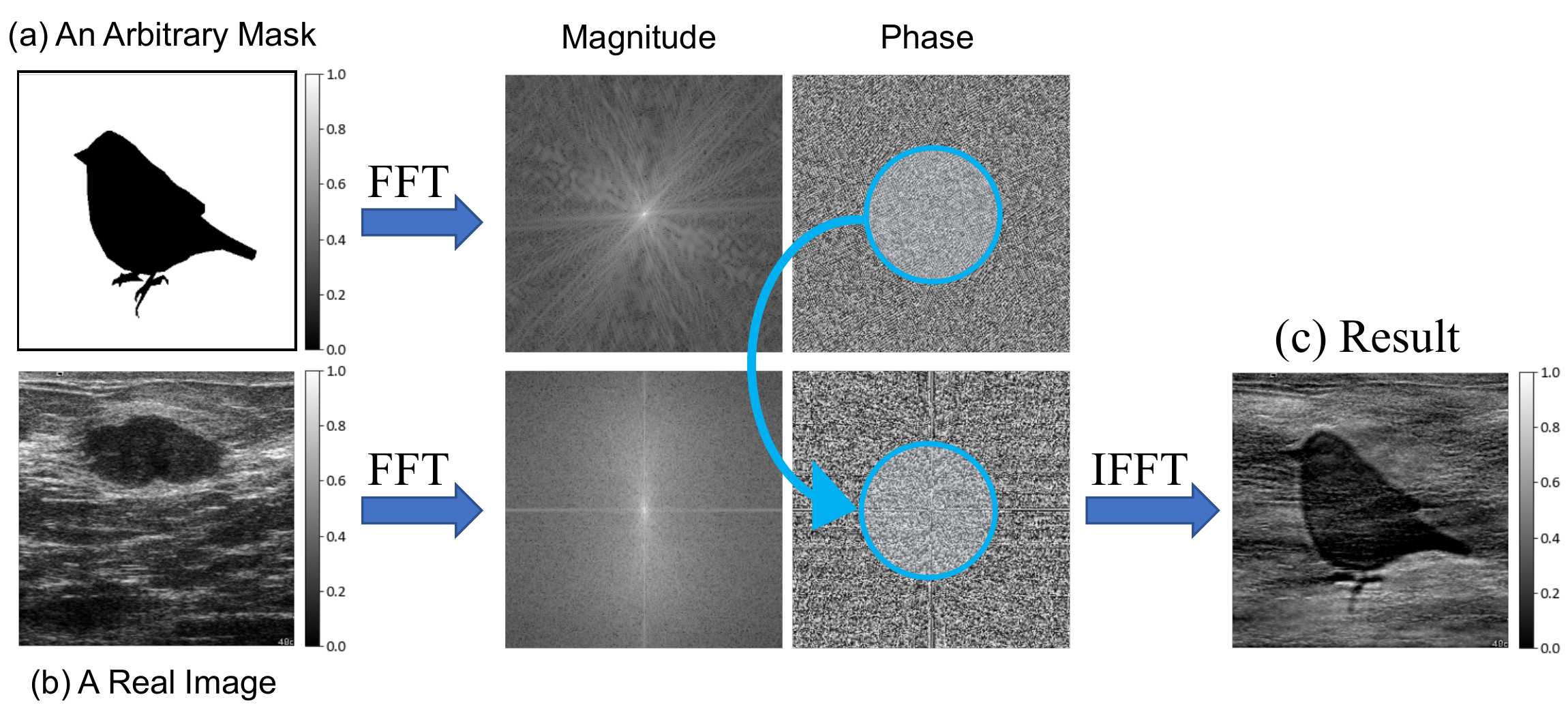}
	\caption{Given an arbitrary mask and a real US image, the proposed method takes the FFT of both inputs and replaces the phase information of the low-frequency spectrum of the real image with the corresponding information of the arbitrary mask to generate the output. (a) An arbitrary mask. (b) A real US image. (c) The simulated output image.}
	\label{fig1}
\end{figure*}

\noindent Finally, given a pair of a real image and an arbitrary mask, the proposed method for simulating a new image can be formulated as:
\begin{flalign}
I_s = \mathcal{F}^{-1}(\mathcal{F}_M(I_r),  M_\alpha\boldsymbol{\cdot}\mathcal{F}_P(I_m)+(1-M_\alpha)\boldsymbol{\cdot}\mathcal{F}_P(I_r))
\label{eq4}
\end{flalign}
\noindent where $\alpha \in \mathbb{R}$ is a parameter that specifies the amount of phase information that needs to be replaced, and in this work, we set $\alpha=0.11$.
Fig. \ref{fig1} illustrates the proposed method, where (a) is an arbitrary mask $I_m$, and (b) is a real US image $I_r$. After taking the FFT of both images, we replaced the phase information of the low-frequency spectrum of the real image with the corresponding information of the mask. Finally, by taking the IFFT of the modified real image $I_r$, the simulated image $I_s$ was generated.

\section{Segmentation Task}
Given a sample input image $I\in \mathbb{R}^{W\times H}$ and its corresponding output segmentation mask $\hat{S}\in \{0,1\}^{W\times H}$,  the segmentation problem can be formulated as:
\begin{flalign}
\hat{S} = f_{seg}(I, \boldsymbol{\theta})
\end{flalign}
\noindent where $W$ and $H$ are width and height of the image, respectively, $f_{seg}: \mathbb{R}^{W\times H} \rightarrow \{0,1\}^{W\times H}$ is the segmentation CNN, and $\boldsymbol{\theta}$ are the network's parameters. By training the model, an optimizer tries to find optimal parameters $\boldsymbol{\theta^{\ast}}$ that minimize the error, measured by a loss function $L$, between predicted mask $\hat{S}$ and ground truth $S$
\begin{flalign}
\boldsymbol{\theta^{\ast}} = \underset{\theta}{argmin} \; L(S, \hat{S})
\end{flalign}

\subsection{Datasets}
\subsubsection{\textit{In vivo} Dataset}
We utilized a publicly available US breast images dataset, known as Dataset B \cite{Yap2018}, which was collected in 2012 from the UDIAT Diagnostic Centre with a Siemens ACUSON Sequoia C512 system and a 17L5 HD linear array transducer with a frequency of 8.5 MHz. The dataset consisted of 163 breasts B-mode US images from different women with a mean image size of 760 × 570 pixels, where each one included lesions of different sizes at different locations. Lesions were categorized into two classes of benign and cancerous, with 110 and 53 images in each class, respectively. Corresponding lesion masks were also delineated by experienced radiologists and provided along with the dataset as ground truth masks. We resampled all images to a size of 256$\times$256 pixels and split the dataset into three training, validation, and test sets, each containing 20, 20, and 123 images, respectively.

\subsubsection{Field II Dataset}
To compare the proposed method with Field II, we simulated 1000 images using this publicly available simulation package \cite{Jensen1992, Jensen1996}. Each image contained 100,000 scatterers uniformly distributed inside a phantom of size 50 mm $\times$ 10 mm $\times$ 50 mm in $x$, $y$, and $z$ directions, respectively. Phantoms were positioned at an axial depth of 20 mm from the face of the transducer and centered at the focal point.  Besides, we added an anechoic region with an arbitrary shape to each one. To generate those anechoic regions, we took 1000 samples with only one salient object from a publicly available dataset, known as XPIE \cite{Xia2017}, which contained segmented natural images. Then we discarded natural images and resampled only their ground truth masks with the same size as the phantom. Finally, we assigned a zero weight to the amplitude of those scatterers which were located inside the mask. The advantages of this method were twofold: First, it enabled us to consider the masks as the ground truths of simulated images. Second, we provided the network with a wider range of features compared to regions with limited shapes. Finally, we resampled all images to 256$\times$256 pixels, and split them into two training and validation sets, each containing 800 and 200 images, respectively. Note that this data was merely used for training and validation and did not contain a test set.
The parameters of the Field II simulation are summarized in Table~\ref{tbl1}.
\begin{table}[h!]
	\caption{Field II parameters for data simulation.}
	\label{tbl1}
	\setlength{\tabcolsep}{3pt}
	\def\arraystretch{1.5}%
	\begin{tabular}{p{115pt}p{100pt}}
		\hline
		\textbf{Parameter}& 
		\textbf{Value}\\
		\hline
		Number of Lines& 
		50\\
		Number of Elements& 
		192\\
		Number of Active Elements& 
		64\\
		Elevation Element Height& 
		5 mm\\
		Element Width& 
		Equals to wavelength\\
		Kerf& 
		0.05 mm\\
		Sound Speed& 
		1540 m/s\\
		\hline
	\end{tabular}
\end{table}

\subsubsection{Ultra-Fast Dataset}
For simulating an image using the proposed method, an arbitrary mask and a real image are required. We took the same 1000 masks from the XPIE dataset, which were used for simulating the Field II dataset, and randomly paired each one with a real image from the training set of the \textit{in vivo} dataset to simulate 1000 new images. Similar to the Field II dataset, we resampled all images to 256$\times$256 pixels and split them into two training and validation sets, each containing 800 and 200 images, respectively, where there was no need for a test set.

\subsection{Network Architecture and Training Strategy}
We used a vanilla U-Net \cite{Ronneberger2015} to evaluate the performance of the proposed method in a segmentation task. U-Net was proposed particularly for biomedical image segmentation, where the size of the training set is small. Its architecture comprises an encoder followed by a decoder, and skip connections are also employed to concatenate low-level features of the encoder with high-level ones in the decoder.

For the training process, we set the learning rate and batch size to $1\times 10^{-4}$ and 16, respectively. The AdamW \cite{Loshchilov2019}, a variant of Adam \cite{Kingma2015}, with a weight decay of $10^{-2}$ was exploited as the optimizer, and a sigmoid activation function was used for the output layer.
The loss function was defined based on the Dice similarity coefficient (DSC). This metric quantifies the area overlap between the ground truth and predicted masks and was also used for evaluating the segmentation performance:
\begin{flalign}
DSC(S,\hat{S}) = \frac{2 \left |S \cap \hat{S} \right |+ \varepsilon}{\left |S \right |+ \left | \hat{S} \right | + \varepsilon}
\end{flalign}
where $\varepsilon$ is a small number that prevents numerical instability for small masks. After each training or fine-tuning epoch, model weights were saved only if the validation loss had been improved, and finally, the best model was used for testing. Experiments were implemented using the PyTorch package \cite{Paszke2019} and run on an NVIDIA TITAN Xp GPU with 12 GB of memory.

\begin{figure}
	\centering
	\includegraphics[width=0.9999\linewidth]{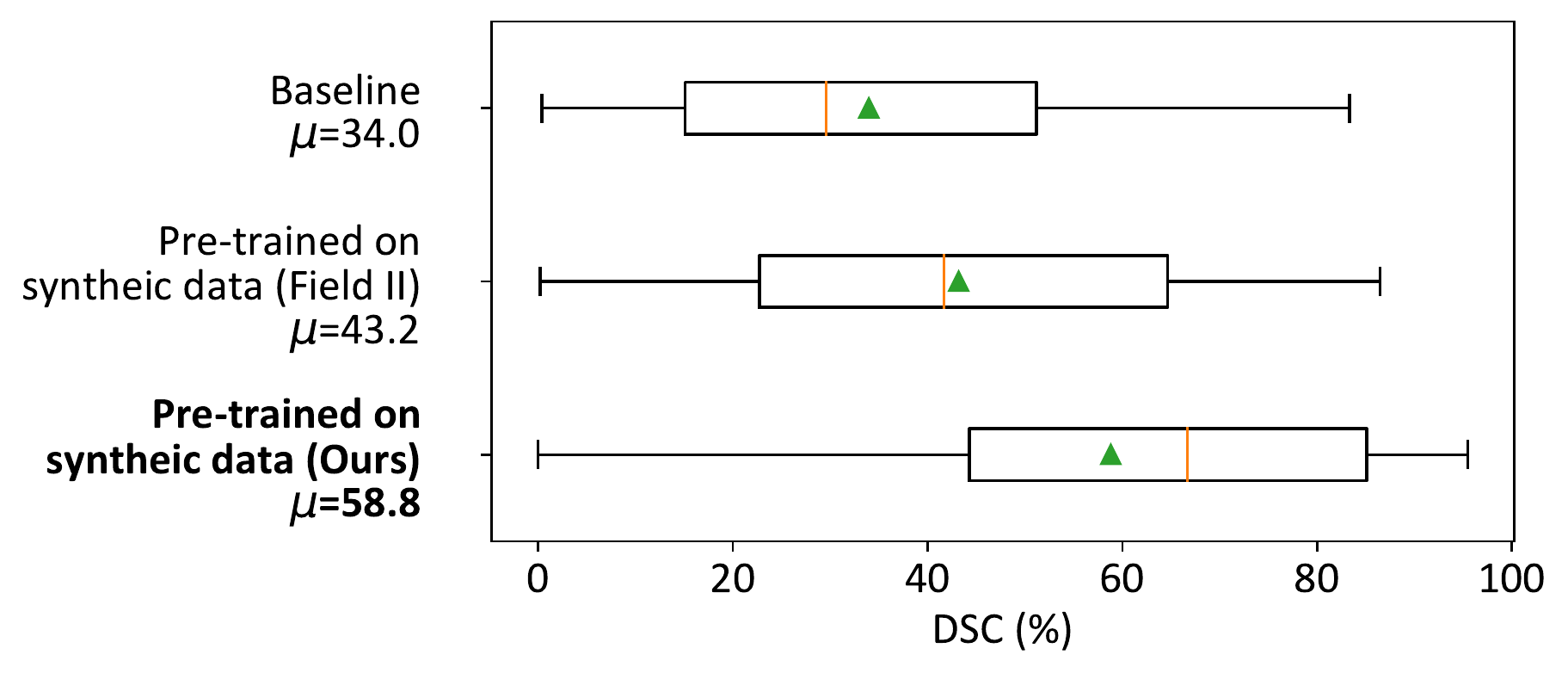}
	\caption{Comparison of DSC over the \textit{in vivo} test set achieved by three conducted experiments. (Top) Training the network from scratch merely using \textit{in vivo} training set. (Middle) Pre-training the network using synthetic data simulated by Field II and then fine-tuning on the \textit{in vivo} training set. (Bottom) Pre-training the network using synthetic data simulated by the ultra-fast proposed method and then fine-tuning on the \textit{in vivo} training set. The triangle and vertical line represent the mean and median, respectively.}
	\label{fig2}
\end{figure}

\section{Results}
To assess the proposed method for improving the performance of the segmentation task, we conducted three separate experiments. In the first one, labeled as the baseline, a network was trained merely using 20 training images of the \textit{in vivo} dataset for 200 epochs.
In the next experiment, first, the network was trained using 800 training images of the Field II dataset for 150 epochs, and then it was fined-tined using 20 training images of the \textit{in vivo} dataset for 50 more epochs.
Finally, as the last experiment, we repeated the second one, except that instead of the Field II dataset, the ultrafast dataset was employed for pre-training the network. As mentioned before, to avoid data leakage, only the training set of the \textit{in vivo} dataset had been used for simulating the ultrafast dataset.

Fig. \ref{fig2} shows the DSC results over the test set of the \textit{in vivo} dataset for all three experiments. As expected, the baseline method achieved the lowest mean DSC due to training on 20 real US images and without pre-training on simulated images. The second experiment obtained a better performance by taking advantage of pre-training on synthetic images simulated by Field II and then fine-tuning on real US images. Finally, the third experiment demonstrated that pre-training the network on synthetic data simulated by the proposed method achieved 24.8\% higher mean DSC than the baseline experiment and outperformed Field II simulations by 15.6\% improvement in mean DSC. Another advantage of the proposed method is that most of its computational cost is devoted to taking FFT and IFFT, which made simulating the ultra-fast dataset almost 36000 times faster than the Field II dataset using the same CPU.

\section{Conclusion}
We introduced a novel ultra-fast approach based on the Fourier transform for simulating US images. In this approach, in contrast with the existing methods such as solving acoustic wave equations, employing ray-tracing, or using GANs, we proposed replacing the phase information of the low-frequency spectrum of a real US image with the corresponding information of an arbitrary mask to simulate a new image containing lesion(s) with known ground truth. We assessed the utility of this method in a lesion segmentation task, where a U-Net was pre-trained using synthetic data. We demonstrated that images simulated by the proposed method outperformed Field II simulations in terms of improving the mean DSC by 15.6\%, while the simulation was almost 36000 times faster (both on CPU).

\section*{Acknowledgment}
The authors would like to thank Natural Sciences and Engineering Research Council of Canada (NSERC) for funding. We thank NVIDIA for donating the GPU.

\bibliographystyle{IEEEtran}
\bibliography{bibliography}

\end{document}